\documentclass[11pt]{article}
\usepackage{amsmath}
\usepackage{amssymb}
\def\xxx#1 {{\sf hep-th/#1} }

\def\G{\Gamma}
\def\D{\Delta}
\def\L{\Lambda}

\def\a{\alpha}
\def\b{\beta}

\def\d{\delta}
\def\e{\varepsilon}

\def\m{\mu}

\def\s{\sigma}
\def\r{\rho}
\def\l{\lambda}
\def\t{\tau}
\def\o{\omega}

\def\vt{\vartheta}
\def\mc{\mathcal}

\def\p{\partial}

\def\ra{\rangle}
\def\dg{\dagger}
\def\wt{\widetilde}
\numberwithin{equation}{section} 
\parskip=\medskipamount

\begin{document}
\thispagestyle{empty} \addtocounter{page}{-1}
\begin{flushright}
\vskip-0.5cm
AEI-2002-064 \\
{\tt hep-th/0208209}
\end{flushright}
\vspace*{2cm} \centerline{\Large \bf  More comments on
superstring interactions in the pp-wave background}
\vspace*{1.5cm} \centerline{Ari Pankiewicz}
\begin{center}
\it Max-Planck-Institut f\"ur Gravitationsphysik, Albert-Einstein-Institut \\
Am M\"uhlenberg 1, D-14476 Golm, \rm GERMANY \\
\vspace*{0.2cm}
\end{center}
\vspace*{0.7cm} \centerline{\tt apankie@aei-potsdam.mpg.de}
\vspace*{1.5cm} \centerline{\bf abstract} \vspace*{0.5cm}
We reconsider light-cone superstring field theory on the maximally
supersymmetric pp-wave background. We find that the results for the fermionic
Neumann matrices given so far in the literature are incomplete and verify our
expressions by relating them to the bosonic Neumann matrices and proving
several non-trivial consistency conditions among them, as for example the
generalization of a flat space factorization theorem for the bosonic Neumann matrices.
We also study the bosonic and fermionic constituents of the prefactor and point out
a subtlety in the relation between continuum and oscillator basis expressions.
\noindent \vspace*{1.1cm}
\baselineskip=18pt
\newpage
\section{Introduction}
As is well known by now, besides flat space and AdS$_5\times S^5$ type IIB superstring
theory admits an additional solution preserving all 32 supersymmetries, the 
maximally supersymmetric pp-wave background \cite{bfhp}. In contrast to flat space and
$AdS_5\times S^5$ which are related being the asymptotic and near-horizon geometry of the
D3-brane respectively, the maximally supersymmetric pp-wave solution is obtained from
$AdS_5\times S^5$ by the Penrose limit \cite{penrose}, i.e.
blowing up the neighborhood of the trajectory of a massless particle rotating around a
great circle of the $S^5$. This observation led Berenstein, Maldacena and Nastase (BMN)
\cite{bmn} to a generalization of the AdS/CFT correspondence to the pp-wave background,
by proposing that on the Yang-Mills side the Penrose limit is mimicked by focusing on a
subset of composite operators of ${\mc N}=4$ SYM with large R charge. The main importance
of this conjectured correspondence lies in the fact that, in contrast to IIB string theory
on AdS$_5\times S^5$ which due to the presence of the RR 5-form flux is still
largely intractable, in the light-cone GS formalism type IIB theory in the maximally
supersymmetric pp-wave background is free and thus exactly solvable \cite{m,mt}
despite of the RR flux.

Due to the solvability of the theory it is natural to try to include interactions in the
picture. In light-cone (closed) superstring field theory interactions are encoded in a cubic
interaction vertex, which in flat space was studied in detail by Green,
Schwarz and Brink \cite{gs,gsb} and recently has been generalized to the pp-wave background
by Spradlin and Volovich \cite{sv,sv2}. The interaction vertex can be split into two parts,
one -- the exponential part of the vertex -- contains the Neumann matrices and is necessary
to impose the kinematic constraints of the interaction \cite{gs,gsb}. The second part
-- the so called prefactor -- implements the dynamical constraint that the superalgebra is
realized in the interacting theory \cite{gs,gsb}.

In the usual AdS/CFT correspondence it is used that $SU(N)$ ${\mc N}=4$ SYM perturbation
theory can be written as a double series expansion in the
genus counting parameter $1/N^2$ and the 't Hooft coupling $\l=g^2_{\text{YM}}N$.
String theory on AdS$_5\times S^5$ has three parameters, the string length $\sqrt{\a'}$,
the string coupling $g_s$ and -- characterizing the background --
the AdS and sphere radius $R$ from which we can form two dimensionless quantities to be
matched with corresponding parameters of the gauge theory
\begin{equation}
\frac{R^2}{\a'}=\sqrt{\l}\,,\qquad\text{and}\qquad 4\pi g_s=g^2_{\text{YM}}\,.
\end{equation}
In particular the supergravity approximation, where explicit
calculations can be performed, corresponds to strongly coupled gauge theory and this makes
a detailed comparison of non-protected quantities very difficult.
In the pp-wave case one considers the sector of operators in the SYM theory with
$U(1)$ charge $J$ and conformal dimension $\D$ such that the difference $\D-J$ remains fixed
in the limit $J$, $\D\to\infty$. One also takes
$N\to\infty$ such that $J\sim\sqrt{N}$ and keeps $g^2_{\text{YM}}$ finite \cite{bmn}.
There is evidence \cite{jan,freedman}
that in this case the theory can again be expanded in a double series in the effective genus
counting parameter $g_2^2=\frac{J^4}{N^2}$ and the effective coupling
$\lambda'=\frac{g^2_{\text{YM}}N}{J^2}$. Moreover the finite
quantity $\D-J$ turns out to be a function of $\lambda'$ and $g_2^2$ \cite{bmn, gross,sz}.
So we are still left with two parameters on the gauge theory side in the pp-wave limit,
although we introduced the quantity $J$ and so $\lambda'$ and $g_2^2$ depend on three
parameters. However, since both $J$ and $N$ are strictly taken to infinity, only the ratio
$J^2/N$ survives as a true parameter. Light-cone string theory in the pp-wave background
effectively has three parameters, the string length $|\a|\equiv\a'|p^+|$,
the string coupling $g_s$ and the scale $\m$ characterizing the background.
The matching in this case is \cite{bmn}
\begin{equation}
\frac{1}{(\m\a)^2}=\lambda'\,,\qquad\text{and}\qquad 4\pi g_s(\m\a)^2=g_2\,.
\end{equation}
An interesting property of the duality between string theory on the pp-wave and the subsector
of composite operators in ${\mc N}=4$ SYM with large R charge is that in this case both
sides of the duality are simultaneously accessible in their perturbative regimes.

A definite proposal which quantity on the field theory side should be matched with
three-string amplitudes was put forward in \cite{freedman} for the free, planar limit
corresponding to $\m=\infty$\,\footnote{Whenever we write $\m=\infty$ or
$\m\to 0$ we actually mean the dimensionless quantity $\m|\a|$.} on the string theory side.
Various checks of this proposal and further studies of cubic interactions in the
pp-wave background and their comparison to field theory were done in
\cite{sv2} and~\cite{chu}--\cite{huang2}. 
So called vector operators were studied in \cite{goy}, whose results suggest that the proposal of 
\cite{freedman} has to be modified for mixed scalar/vector operators, in order to match string theory 
predictions.\footnote{I would like to thank Marcus Spradlin for emphasizing this point to me.}
An alternative resolution --  to use a different fermionic zero-mode vertex as compared to \cite{sv} -- 
was proposed in \cite{chu1}. 
The structure of the interaction vertex for large $\m$, in particular subleading
corrections which are important in trying to generalize the proposal of \cite{freedman}
to the interacting gauge theory were studied in \cite{ksv,sch}. The latter has
some overlap with results presented in section \ref{sec4} of this paper. Additional work
on string interactions can be found in \cite{gopakumar,verlinde}, an interesting new
development concerning string theory on non-maximally supersymmetric plane wave backgrounds
is \cite{mm}. For the covariant formulation of string theory on plane wave backgrounds see
\cite{berkovits,bm}. A first study of four-point correlation functions in the BMN limit
was initiated in \cite{jan2}.

This paper is organized as follows. We begin with a
brief review of the free light-cone string theory in section \ref{sec2}. In section
\ref{sec3} we discuss the interacting theory. The first part of this section deals with the
bosonic contribution to the exponential part of the vertex and is included for completeness,
all of the presented formulae were already derived in \cite{sv}. The second part of section
\ref{sec3} deals with the fermionic contribution to the exponential part of the vertex and
slightly differs from the original result of \cite{sv}. A detailed proof of the expressions presented
there is relegated to appendix B. Section \ref{sec4}
deals with properties of the Neumann matrices and the flat space limit of the exponential
part of the vertex, in particular we generalize a flat space factorization theorem \cite{gs}
to the pp-wave background, see also \cite{sch} were the same result is obtained. In section
\ref{sec5} we study in some detail the constituents of the prefactor. We begin with the
oscillator expressions of the bosonic part which were already obtained in \cite{sv2}.
Using results of section \ref{sec4} we then verify explicitly that the operator expressions
\cite{gs,gsb} for the bosonic
constituents coincide up to a possibly $\m$-dependent normalization with the oscillator
expressions when acting on the vertex. In the second part of section \ref{sec5} we
consider the fermionic constituent and relate the vector appearing in the oscillator
expression to its bosonic counterpart and check that it has the correct flat space limit
of \cite{gsb}. In the end of this section we prove that in the fermionic case a non-trivial
$\m$-dependent matrix appears in the relation between operator and oscillator expressions.
We finish with conclusions in section \ref{sec6}, summarize some known identities
\cite{gs} in appendix A and present the alternative solution of the kinematic constraints 
(see~\cite{chu2}) based on the proposal of~\cite{chu1} in appendix~\ref{app:c}.

\section{Review of free string theory on the pp-wave}\label{sec2}
In this section we review some known facts about free light-cone string
field theory in the maximally supersymmetric pp-wave background \cite{m,mt,sv}.
We begin with the bosonic part of the string light-cone action in
the pp-wave background \cite{m,mt}
\begin{equation}
S_{\text{l.c.}}=\frac{e(\a)}{4\pi\a'}\int\,d\t\int_0^{2\pi|\a|}\,d\s\bigl[\dot{x}^2-
\grave{x}^2-\m^2x^2\bigr]\,,
\end{equation}
where $\dot{x}=\p_{\t}x$, $\grave{x}=\p_{\s}x$, $|\a|\equiv\a'|p^+|$,
$e(\a)\equiv\text{sign}(\a)$ and $p^+<0$ ($p^+>0$) for incoming (outgoing) strings. We
suppress the transverse index. The mode expansions of the fields $x(\s,\t)$
and $p(\s,\t)$ at $\t=0$ are
\begin{equation}
\begin{split}
x(\s)& = x_0+\sqrt{2}\sum_{n=1}^{\infty}
\bigl(x_n\cos\frac{n\s}{|\a|}+x_{-n}\sin\frac{n\s}{|\a|}\bigr)\,,\\
p(\s) & =\frac{1}{2\pi|\a|}\bigl[p_0+\sqrt{2}\sum_{n=1}^{\infty}
\bigl(p_n\cos\frac{n\s}{|\a|}+p_{-n}\sin\frac{n\s}{|\a|}\bigr)\bigr]\,,
\end{split}
\end{equation}
where in terms of oscillators
\begin{equation}\label{xp}
x_n=i\sqrt{\frac{\a'}{2\o_n}}\bigl(a_n-a_n^{\dg}\bigr)\,,\qquad
p_n=\sqrt{\frac{\o_n}{2\a'}}\bigl(a_n+a_n^{\dg}\bigr)\,,\qquad
[a_n,a_m^{\dg}]=\d_{nm}
\end{equation}
and $\o_n=\sqrt{n^2+(\m\a)^2}$. The light-cone hamiltonian is
\begin{equation}
H=\frac{1}{\a}\sum_{n\in{\bf Z}}\o_n\bigl(a_n^{\dg}a_n+4\bigr)\,.
\end{equation}
The zero-point energy will be cancelled by the fermionic
contribution. The hamiltonian only depends on the two dimensionful
quantities $\m$ and $\a$, i.e. $\a'$ and $p^+$ should not be
treated separately.

The fermionic part of the action is \cite{m,mt}
\begin{equation}
S_{\text{l.c.}}=\frac{1}{8\pi}\int\,d\t\int_0^{2\pi|\a|}\,d\s[i(\bar{\vt}\dot{\vt}+\vt\dot{\bar{\vt}})
-\vt\grave{\vt}+\bar{\vt}\grave{\bar{\vt}}-2\m\bar{\vt}\Pi\vt]\,,
\end{equation}
where $\vt^a$ is a complex positive chirality spinor of $SO(8)$
(we mostly suppress the index $a$)
and $\Pi=\G^1\G^2\G^3\G^4$ is symmetric and traceless, $\Pi^2=1$.

The mode expansion of $\vt$ and its conjugate momentum
$i\l\equiv\frac{i}{4\pi}\bar{\vt}$ at $\t=0$ is
\begin{equation}
\begin{split}
\vt(\s) & =\vt_0+\sqrt{2}\sum_{n=1}^{\infty}
\bigl(\vt_n\cos\frac{n\s}{|\a|}+\vt_{-n}\sin\frac{n\s}{|\a|}\bigr)\,,\\
\l(\s) & =\frac{1}{2\pi|\a|}\bigl[\l_0+\sqrt{2}\sum_{n=1}^{\infty}
\bigl(\l_n\cos\frac{n\s}{|\a|}+\l_{-n}\sin\frac{n\s}{|\a|}\bigr)\bigr]
\end{split}
\end{equation}
with the reality condition $\l_n=\frac{|\a|}{2}\bar{\vt}_{n}$.
The anticommutation relation
$\{\vt^a(\s),\l^b(\s')\}=\d^{ab}\d(\s-\s')$ follows from
$\{\vt^a_n,\l^b_m\}=\d^{ab}\d_{n,m}$. To write the hamiltonian in canonical form in terms of
fermionic operators $b_n$ satisfying $\{b_n,b_m^{\dg}\}=\d_{nm}$ define \cite{sv}
\begin{equation}
\vt_n=\frac{c_n}{\sqrt{|\a|}}\left[(1+\r_n\Pi)b_n+e(\a
n)(1-\r_n\Pi)b_{-n}^{\dg}\right]\,,
\end{equation}
where $e(n)=\text{sign}(n)$, ($e(0)\equiv 1$) and
\begin{equation}
\r_n=\r_{-n}=\frac{\o_n-|n|}{\m\a}\,,\qquad
c_n=c_{-n}=\frac{1}{\sqrt{1+\r_n^2}}\,.
\end{equation}
Then the fermionic part of the light-cone hamiltonian is
\begin{equation}
H=\frac{1}{\a}\sum_{n\in{\bf Z}}\o_n\bigl(b_n^{\dg}b_n-4\bigr)\,.
\end{equation}
Notice that in the limit $\m\to0$ we have $c_n(1\pm\r_n\Pi)\to 1$ and
therefore
\begin{equation}
\lim_{\m\to0}\vt_n=\frac{1}{\sqrt{|\a|}}\bigl(b_n+e(\a
n)b_{-n}^{\dg}\bigr)\,.
\end{equation}
This allows to relate the $b_n$ to the $Q_n^{I\,,\,II}$ of
\cite{gsb} which will be useful when checking that the fermionic
vertex has the correct flat space limit. For $n>0$ we have in the
limit $\m\to0$
\begin{equation}\label{oscrel}
b_n \longleftrightarrow \frac{e(\a)}{\sqrt{|\a|}}Q_n^I\,,\qquad
b_{-n} \longleftrightarrow \frac{e(\a)}{\sqrt{|\a|}}Q_n^{II}\,,
\end{equation}
and $\left[Q_n^{I\,,\,II}\right]^{\dg}=e(\a)Q_{-n}^{I\,,\,II}$.
\section{Interacting string theory}\label{sec3}
As already stated in the introduction, the cubic interaction vertex of light-cone
string field theory can be split into two parts, determined by imposing the
kinematic and dynamic constraints of the interaction. In this section
we study the exponential part of the vertex which deals with the kinematic constraints.
The prefactor that is determined by the dynamic constraints is the subject of section
\ref{sec5}. The bosonic contribution to the exponential part of the three-string interaction
vertex has to satisfy the kinematic constraints \cite{gs,gsb}
\begin{equation}
\sum_{r=1}^3p_r(\s_r)|E_a\ra=0\,,\qquad
\sum_{r=1}^3e(\a_r)x_r(\s_r)|E_a\ra=0\,.
\end{equation}
It can be obtained by evaluating the integral \cite{gs,gsb}
\begin{equation}\label{ea}
|E_a\ra=\prod_{r=1}^3\int\,dp_r\psi(p_r)\D^8\bigl[\sum_{s=1}^3p_s(\s_s)\bigr]|0\ra=
\prod_{r=1}^3\prod_{n\in{\bf Z}}\int\,dp_{n(r)}\psi(p_{n(r)})
\d^8\bigl[\sum_{s=1}^3\bigl(X^{(s)}p_s\bigr)_n\bigr]|0\ra\,.
\end{equation}
$\D^8\bigl[\sum_{s=1}^3p_s(\s_s)\bigr]$ is the Delta-functional guaranteeing the
continuous overlap of the string worldsheets in the interaction and
$\psi(p_{n(r)})$ is the harmonic oscillator wavefunction for
occupation number $n$
\begin{equation}\label{boswv}
\psi(p_{n(r)})=\Bigl(\frac{\o_{n(r)}\pi}{\a'}\Bigr)^{-1/4}
\exp\left(-\frac{\a'}{(2\o_{n(r)})}p_{n(r)}^2+
\sqrt{\frac{2\a'}{\o_{n(r)}}}a_{n(r)}^{\dg}p_{n(r)}
-\frac{1}{2}a_{n(r)}^{\dg}a_{n(r)}^{\dg}\right)\,.
\end{equation}
The coordinates of the three strings are parameterized by
\begin{align}
\s_1 & =\s \qquad\quad\qquad -\pi\a_1\le\s\le\pi\a_1\,, \nonumber \\
\s_2 & =\begin{cases} \s-\pi\a_1 & \quad\pi\a_1\le\s\le\pi(\a_1+\a_2)\,, \\
\s+\pi\a_1 & \quad-\pi(\a_1+\a_2)\le\s\le-\pi\a_1\,, \end{cases} \\
\s_3 & =-\s \qquad\quad\qquad -\pi(\a_1+\a_2)\le\s\le
\pi(\a_1+\a_2)\nonumber\,.
\end{align}
Here $\a_1+\a_2+\a_3=0$ and $\a_3<0$. We also use
$\b\equiv\a_1/\a_3$ and $\b+1=-\a_2/\a_3$.
The full Delta-functional takes the form \cite{gs,gsb}
\begin{equation}
\D\bigl[\sum_{r=1}^3p_r(\s_r)\bigr]\sim\prod_{m\in{\bf Z}}
\d\left(\sum_{r=1}^3\sum_{n\in{\bf
Z}}X^{(r)}_{mn}p_{n(r)}\right)\,.
\end{equation}
We ignored factors of $\sqrt{2}$ which can be absorbed in the
measure. The matrices $X_{mn}^{(r)}$ have the following structure\cite{gs,gsb}\footnote{The
matrices $X^{(r)}_{mn}$ actually differ from the ones of \cite{gs,gsb}
by a factor of $(-1)^m$, which however has no physical significance.}
\begin{equation}
X^{(r)}_{mn}\equiv\begin{cases}
X^{(r)}_{mn}\,,\qquad m>0\,,n>0 \\ \frac{\a_3}{\a_r}\frac{n}{m}X^{(r)}_{-m,-n}\,,\qquad m<0\,,n<0 \\
\frac{1}{\sqrt{2}}X^{(r)}_{m0}\,,\qquad m>0 \\ 1\,,\qquad m=0=n \\
0\,,\qquad\text{otherwise}\,.
\end{cases}
\end{equation}
Here \cite{gs,gsb}
\begin{equation}
X^{(1)}_{mn}\equiv(-1)^n\frac{2m\b}{\pi}\frac{\sin m\pi\b}{m^2\b^2-n^2}\,,\qquad
X^{(2)}_{mn}\equiv\frac{2m(\b+1)}{\pi}\frac{\sin m\pi\b}{m^2(\b+1)^2-n^2}
\end{equation}
and $X^{(3)}_{mn}=\d_{mn}$.
It is standard to perform the gaussian integral \eqref{ea} and the result is \cite{gs,gsb,sv}
\begin{equation}\label{bv}
|E_a\ra\sim\exp\left(\frac{1}{2}a^{\dg
T}_r\overline{N}^{rs}a^{\dg}_s\right)|0\ra\,,
\end{equation}
where $r,s\in\{1,2,3\}$.
The determinant factor coming from the
functional determinants will be cancelled by the fermionic
contribution except for the zero-mode part which is proportional
to $(\frac{\m}{\a'})^2(\frac{\a_1\a_2}{\a_3})^2$.

The Neumann matrices are \cite{sv}
\begin{equation}
\overline{N}^{rs}_{mn}=\d^{rs}\d_{mn}-2\bigl(C_{(r)}^{1/2}X^{(r)T}\G_a^{-1}X^{(s)}C_{(s)}^{1/2}\bigr)_{mn}\,,
\end{equation}
where
\begin{equation}
\bigl[C_{(r)}\bigr]_{mn}=\o_{m(r)}\d_{mn}\,,\qquad\text{and}\qquad
\G_a=\sum_{r=1}^3 X^{(r)}C_{(r)}X^{(r)T}\,.
\end{equation}
From the structure of the $X^{(r)}$ it follows that $\G_a$ is
block diagonal and using the identities \eqref{id2} in
appendix A one can write the blocks as \cite{sv} ($C_{mn}=m\d_{mn}$)
\begin{equation}
\bigl[\G_a\bigr]_{mn}=
\begin{cases}
\bigl(C^{1/2}\G C^{1/2}\bigr)_{mn}\,, & m,n>0 \,, \\
-2\m\a_3\,, & m=0=n\,, \\
\bigl(C^{1/2}\G_-C^{1/2}\bigr)_{-m,-n}\,, & m,n<0\,,
\end{cases}
\end{equation}
where
\begin{equation}
\G\equiv\sum_{r=1}^3 A^{(r)}U_{(r)}A^{(r)\,T}\,,\qquad
\G_-\equiv\sum_{r=1}^3 A_-^{(r)}U^{-1}_{(r)}A_-^{(r)\,T}\,.
\end{equation}
Here $A^{(r)}_{mn}=\bigl[C^{-1/2}X^{(r)}C^{1/2}\bigr]_{mn}$ for
$m$, $n>0$ and
\begin{equation}
U_{(r)}=C^{-1}\bigl(C_{(r)}-\m\a_r{\bf 1}\bigr)\,,\qquad
U^{-1}_{(r)}=C^{-1}\bigl(C_{(r)}+\m\a_r{\bf 1}\bigr)\,,\qquad
A_-^{(r)}=\frac{\a_3}{\a_r}C^{-1}A^{(r)}C\,.
\end{equation}
The matrix $\G$ (which reduces to the flat space $\G$ of
\cite{gs,gsb} for $\m\to0$) exists and is invertible, whereas
$\G_-$ is ill-defined since the above sum is divergent. Nevertheless it is
possible to derive a well-defined identity for $\G_-^{-1}$ \cite{sv}
\begin{equation}\label{gamma-}
\G_-^{-1}=U_{(3)}\bigl(1-\G^{-1}U_{(3)}\bigr)\,.
\end{equation}
Since $\G_-^{-1}$ is related to $\G^{-1}$ it is possible to relate
the Neumann matrices with positive and negative indices. The only nonvanishing
matrix elements with negative indices are
$\overline{N}^{rs}_{-m,-n}$ for $m$, $n>0$ related to
$\overline{N}^{rs}_{mn}$ via \cite{sv}
\begin{equation}\label{neg}
\overline{N}^{rs}_{-m,-n}=-\left(U_{(r)}\overline{N}^{rs}U_{(s)}\right)_{mn}\,.
\end{equation}

Analogously to the bosonic contribution, the fermionic exponential part of the interaction
vertex has to satisfy \cite{gs,gsb}
\begin{equation}\label{fermcon}
\sum_{r=1}^3\l_{(r)}(\s_r)|E_b\ra=0\,,\qquad
\sum_{r=1}^3e(\a_r)\vt_{(r)}(\s_r)|E_b\ra=0\,.
\end{equation}
As in the bosonic case it could be obtained by constructing the fermionic analogue
of the wavefunction \eqref{boswv} and then performing the resulting integrals
over the non-zero-modes. The pure zero-mode contribution has to be treated separately.
Notice that due to the structure of the $X_{mn}^{(r)}$ the exponential will -- as in flat
space \cite{gs,gsb} -- contain a part which is linear in zero-mode oscillators.
Instead of directly performing the functional integral the exponential can be
obtained (up to the normalization) by making a suitable ansatz and imposing the constraints
\eqref{fermcon} \cite{gs,gsb}. We find the following expression
(cf. appendix B for the details; the notation is defined below)
\begin{equation}\label{fv}
|E_b\ra \sim
\exp\left[\sum_{r,s=1}^3\sum_{m,n=1}^{\infty}b^{\dg}_{-m(r)}Q^{rs}_{mn}b^{\dg}_{n(s)}
-\sqrt{2}\L\sum_{r=1}^3\sum_{m=1}^{\infty}Q^r_mb^{\dg}_{-m(r)}\right]|E_b^0\ra\,,
\end{equation}
where $|E_b^0\ra$ is the pure zero-mode part of the fermionic vertex (see also the
discussion below and apendix C)
\begin{equation}\label{zero}
|E^0_b\ra=\prod_{a=1}^8\left[\sum_{r=1}^3\l_{0(r)}^a\right]|0\ra
\end{equation}
and manifestly satisfies $\sum_{r=1}^3\l_{0(r)}|E^0_b\ra=0$ and
$\sum_{r=1}^3\a_r\vt_{0(r)}|E^0_b\ra=0$. Notice that $|0\ra$ is
not the vacuum defined to be annihilated by the $b_{0(r)}$. Rather
it satisfies $\vt_{0(r)}|0\ra=0$ and $H_{(r)}|0\ra=4\m e(\a_r)|0\ra$
so that the zero should be thought of as the occupation number. In the limit $\m\to 0$
it coincides with the usual flat space vacuum. Furthermore
\begin{align}
\label{qmn}
Q^{rs}_{mn} & =e(\a_r)\sqrt{\left|\frac{\a_s}{\a_r}\right|}
\bigl[P_{(r)}^{-1}U_{(r)}C^{1/2}\overline{N}^{rs}C^{-1/2}U_{(s)}P_{(s)}^{-1}\bigr]_{mn}\,,\\
\label{qm}
Q^r_n & =\frac{e(\a_r)}{\sqrt{|\a_r|}}(1-4\m\a
K)^{-1}(1-2\m\a K(1+\Pi))\bigl[P_{(r)}C_{(r)}^{1/2}C^{1/2}\overline{N}^r\bigr]_n\,,\\
\L & = \a_1\l_{0(2)}-\a_2\l_{0(1)}\,.
\end{align}
Here we introduced some more notation, namely $\a\equiv\a_1\a_2\a_3$,
\begin{equation}
P_{n(r)}\equiv\frac{1-\r_{n(r)}\Pi}{\sqrt{1-\r^2_{n(r)}}}\,,\qquad
K\equiv-\frac{1}{4}B^T\G^{-1}B\,,\qquad
\overline{N}^r\equiv -C^{-1/2}A^{(r)\,T}\G^{-1}B
\end{equation}
and the vector $B_m$ is related to $X^{(r)}_{m0}$ via
$X^{(r)}_{m0}=-\e^{rs}\a_s\bigl(C^{1/2}B\bigr)_m$.
The scalar $K$ and the vector $\overline{N}^r$ reduce to the
quantities defined in \cite{gs,gsb} in the flat space limit.
Notice that our result \eqref{fv} is slightly different from the one
obtained in \cite{sv}, in particular the terms proportional to
$\L$ were absent.  We will show in section \ref{sec4} that as
$\m\to0$ our expression \eqref{fv} coincides with the flat space
result of \cite{gsb}. This however does not really prove that the above expression
is correct since some structure is lost in the flat space limit. The analysis of
appendix B however uniquely fixes the above expressions. 
Another important point is the
following: in a recent paper~\cite{chu1} it was proposed that in
order to resolve present discrepancies~\cite{goy} with the field
theory proposal of~\cite{freedman} one should use a different
zero-mode vertex $|\d\ra$ built on the vacuum $|v\ra$ annihilated by all the
$b_{0(r)}$. In particular it was shown that the states $|v\ra$ and $|0\ra$ have 
opposite parity with respect to the discrete $\mathbb{Z}_2$ symmetry that exchanges the 
two transverse $\mathbb{R}^4$'s of the background~\cite{chu1}. To preserve the $\mathbb{Z}_2$ symmetry
in the interacting theory one has to assign positive parity to $|0\ra$ if one uses the zero-mode vertex 
in equation~(\ref{zero}), whereas the proposal of~\cite{chu1} requires positive parity for $|v\ra$. Both
assignments seem plausible: in the limit $\m\to 0$ $|0\ra$ reduces to the state in flat space which is 
a $SO(8)$ scalar, so continuity suggests that $|0\ra$ should have positive parity; on the other hand as $|v\ra$ is the true vacuum 
in the pp-wave background it might be more natural to expect that it has positive parity. It was shown in~\cite{chu2} that it is 
possible to extend their proposal to a solution of the full kinematic constraints~(\ref{fermcon}), see also appendix~\ref{app:c}. 
Clearly it is necessary and interesting to try to extend this to the level of the dynamical constraints. 
\section{Properties of the Neumann matrices and the flat space limit}\label{sec4}
In this section we analyze in more detail the exponential part of the vertex presented
in the previous section. We check that in the flat space limit all the expressions
reduce to the known ones derived long ago by Green, Schwarz and Brink \cite{gsb}.
As already stated this is only a consistency check which
nevertheless is illuminating. A detailed proof of the expressions appearing in the exponential part of the
vertex is presented in appendix B. Furthermore we generalize a flat space
factorization theorem \cite{gs} for the bosonic Neumann matrices to the pp-wave
background (the same expression was obtained independently in \cite{sch}) which
might be useful for various purposes, such as the comparison of string and field theory
computations.

For $m,n>0$ the nonvanishing elements of the bosonic Neumann
matrices are \cite{sv,sv2}
\begin{align}
\label{mn} \overline{N}^{rs}_{mn} & =\d^{rs}\d_{mn}
-2\sqrt{\frac{\o_{m(r)}\o_{n(s)}}{mn}}\left(A^{(r)\,T}\G^{-1}A^{(s)}\right)_{mn}\,,\\
\label{m0} \overline{N}^{rs}_{m0} & =
-\sqrt{2\m\a_s\o_{m(r)}}\e^{st}\a_t\overline{N}^r_m\,,\qquad s\in\{1,2\}\,,\\
\label{00a} \overline{N}^{rs}_{00} & = \left(\d^{rs}+\frac{\sqrt{\a_r\a_s}}{\a_3}\right)
\left(1-4\m\a K\right)\,,
\qquad r,s\in\{1,2\}\,,\\
\label{00b} \overline{N}^{r3}_{00} & =
\d^{r3}-\sqrt{\left|\frac{\a_r}{\a_3}\right|}\,,\qquad
r\in\{1,2\}\,.
\end{align}
The terms in $\overline{N}^{rs}_{00}$
and $\overline{N}^{r3}_{00}$ that are not proportional to $\m$
give the pure supergravity contribution to the Neumann matrices.
The part of $\overline{N}^{rs}_{00}$ that is proportional to $\m$
is induced by positive string modes of $p_3$. The only
nonvanishing matrix elements with negative indices are
$\overline{N}^{rs}_{-m,-n}$ related to the Neumann matrices with
positive indices by \eqref{neg}.
In flat space $\overline{N}^{rs}_{mn}$ is related to
$\overline{N}^r_m\overline{N}^s_n$ via\footnote{Notice that in
comparison with \cite{gs} we have
$\overline{N}^{rs}_{\text{here}}=C^{1/2}\overline{N}^{rs}_{\text{there}}C^{1/2}$.}
\cite{gs}
\begin{equation}\label{nnflat}
\overline{N}^{rs}_{mn}=-\a\frac{(mn)^{3/2}}{\a_rn+\a_sm}\overline{N}^r_m\overline{N}^s_n\,.
\end{equation}
Below we will derive a generalization of this formula for all
$\m$ (see also \cite{sch}).
Let us introduce \\
$\Upsilon\equiv\sum_{r=1}^3A^{(r)}U_{(r)}^{-1}A^{(r)\,T}=
\G+\m\a BB^T$ (cf. \eqref{id2}).
Its inverse is related to $\G^{-1}$ by (see also \cite{sch})
\begin{equation}\label{ups1}
\Upsilon^{-1}=\G^{-1}-\frac{\m\a}{1-4\m\a K}\left(\G^{-1}B\right)\left(\G^{-1}B\right)^T
\end{equation}
and thus
\begin{equation}\label{ups2}
\Upsilon^{-1}B=\frac{1}{1-4\m\a K}\G^{-1}B\,.
\end{equation}
Then one can show that the following relations hold
\begin{align}
A^{(r)\,T}C^{-1}U_{(3)}\G^{-1} & =
A^{(r)\,T}C^{-1}+\frac{\a_r}{\a_3}C^{-1}U_{(r)}A^{(r)\,T}\G^{-1}\,,\qquad r\in\{1,2\}\,,\\
\Upsilon^{-1}U_{(3)}^{-1}C^{-1}A^{(r)} & =
C^{-1}A^{(r)}+\frac{\a_r}{\a_3}\Upsilon^{-1}A^{(r)}U_{(r)}^{-1}C^{-1}\,,
\qquad r\in\{1,2\}\,,\\
2C^{-1} & = \G^{-1}U_{(3)}C^{-1}+C^{-1}U_{(3)}\G^{-1}+ \Upsilon^{-1}U_{(3)}^{-1}C^{-1}
+C^{-1}U_{(3)}^{-1}\Upsilon^{-1}\nonumber\\
&-\a_1\a_2\Upsilon^{-1}B\left(\G^{-1}B\right)^T\,.
\end{align}
From here we find using \eqref{ups1}, \eqref{ups2} and \eqref{id1}
\begin{align}\label{nnpp}
\overline{N}^{rs}_{mn} & =-(1-4\m\a
K)^{-1}\frac{\a}{\a_r\o_{n(s)}+\a_s\o_{m(r)}}
\left[U_{(r)}^{-1}C_{(r)}^{1/2}C\overline{N}^r\right]_m\left[U_{(s)}^{-1}C_{(s)}^{1/2}C\overline{N}^s\right]_n\,.
\end{align}
Clearly equation \eqref{nnpp} reduces to equation \eqref{nnflat}
as $\m\to0$. It coincides with the result obtained by \cite{sch}.
It is useful to write the above in matrix form as
\begin{equation}\label{nnpp2}
\a_sC_{(r)}\overline{N}^{rs}+\a_r\overline{N}^{rs}C_{(s)}=-\a(1-4\m\a
K)^{-1}
U_{(r)}^{-1}C_{(r)}^{1/2}C\overline{N}^r\bigl[U_{(s)}^{-1}C_{(s)}^{1/2}C\overline{N}^s\bigr]^T\,.
\end{equation}

Using the above we can also give a simple expression for
$Q^{rs}_{mn}$
\begin{equation}
Q^{rs}_{mn}=-e(\a_r)\sqrt{\left|\frac{\a_s}{\a_r}\right|\frac{m}{n}}(1-4\m\a
K)^{-1}\frac{\a}{\a_r\o_{n(s)}+\a_s\o_{m(r)}}
\bigl[P_{(r)}^{-1}C_{(r)}^{1/2}C\overline{N}^r\bigr]_m\bigl[P_{(s)}^{-1}C_{(s)}^{1/2}C\overline{N}^s\bigr]_n\,.
\end{equation}

In what follows we will show that as $\m\to0$ the expressions for
$|E_a\ra$ and $|E_b\ra$ (cf. \eqref{bv} and \eqref{fv}) coincide with the
flat space expressions of \cite{gsb}. We begin with the bosonic
contribution. In the limit $\m\to0$ and for $n>0$ the $a_n$ are related to the
$a_n^{I\,,\,II}$ of \cite{gsb} as
\begin{equation}\label{oscid1}
a_n \longleftrightarrow \frac{1}{\sqrt{n}}a_n^I\,,\qquad a_{-n}
\longleftrightarrow \frac{i}{\sqrt{n}}a_n^{II}
\end{equation}
and $\left[a_n^{I\,,\,II}\right]^{\dg}=a_{-n}^{I\,,\,II}$. Rewrite
\begin{align}
&\frac{1}{2}\sum_{r,s}\sum_{m,n\in{\bf
Z}}a^{\dg}_{m(r)}\overline{N}^{rs}_{mn}a^{\dg}_{n(s)}=
\nonumber\\
&\frac{1}{2}\sum_{r,s}\left[\sum_{m,n=1}^{\infty}a^{\dg}_{m(r)}\overline{N}^{rs}_{mn}a^{\dg}_{n(s)}
+\sum_{m,n=1}^{\infty}a^{\dg}_{-m(r)}\overline{N}^{rs}_{-m,-n}a^{\dg}_{-n(s)}
+2\sum_{m=1}^{\infty}a^{\dg}_{m(r)}\overline{N}^{rs}_{m0}a^{\dg}_{0(s)}+
a^{\dg}_{0(r)}\overline{N}^{rs}_{00}a^{\dg}_{0(s)}\right]\,.
\end{align}
Consider first the part linear in non-zero-mode oscillators. We
have
\begin{equation}
\sum_{r,s=1}^3\sum_{m=1}^{\infty}a^{\dg}_{m(r)}\overline{N}^{rs}_{m0}a^{\dg}_{0(s)}=
\sqrt{2\m}\sum_{r=1}^3\sum_{m=1}^{\infty}\sqrt{\o_{m(r)}}a^{\dg}_{m(r)}\overline{N}^r_m
\bigl(\a_1\sqrt{\a_2}a^{\dg}_{0(2)}-\a_2\sqrt{\a_1}a^{\dg}_{0(1)}\bigr)\,.
\end{equation}
Using
\begin{equation}
a_0^{\dg}=\sqrt{\frac{\a'}{2\m|\a|}}\,p_0+i\sqrt{\frac{\m|\a|}{2\a'}}\,x_0
\end{equation}
this can be further written as
\begin{equation}
\sqrt{\a'}\sum_{r=1}^3\sum_{m=1}^{\infty}\sqrt{\o_{m(r)}}a^{\dg}_{m(r)}\overline{N}^r_m
\left(\mathbb{P}-i\frac{\m\a}{\a'}\mathbb{R}\right)\,.
\end{equation}
Here $\mathbb{P}=\a_1p_{0(2)}-\a_2p_{0(1)}$ and
$\a_3\mathbb{R}=x_{0(1)}-x_{0(2)}$, $[\mathbb{R},\mathbb{P}]=i$.
The above expression reduces to the one in flat space
\cite{gsb} for $\m\to0$. The part quadratic in non zero-mode oscillators
obviously has the correct flat space limit using \eqref{neg}.

Finally the part quadratic in zero-mode oscillators is
\begin{equation}
\sum_{r,s=1}^3a^{\dg}_{0(r)}\overline{N}^{rs}_{00}a^{\dg}_{0(s)}=
\sum_{r,s=1}^3a^{\dg}_{0(r)}M^{rs}_{\text{Sugra}}a^{\dg}_{0(s)}
+2K\a'\left(\mathbb{P}-i\frac{\m\a}{\a'}\mathbb{R}\right)^2\,.
\end{equation}
In the limit $\m\to0$ the second term reduces to the result in
flat space \cite{gsb}. The first term is divergent in the limit which is due
to the fact that the supergravity particles are no longer confined
by the harmonic oscillator potential and propagate in infinite volume.

Now consider the flat space limit of the fermionic expression
$|E_b\ra$ in \eqref{fv}. We will need that
$\lim\limits_{\m\to0}P_{n(r)}^{-1}=1$. Then we see from equations
\eqref{qmn} and \eqref{qm} that
\begin{equation}
\lim_{\m\to0}Q_{mn}^{rs}=\frac{\sqrt{|\a_r\a_s|}}{\a_r}\bigl[C^{1/2}\overline{N}^{rs}C^{-1/2}\bigr]_{mn}\,,\qquad
\lim_{\m\to0}Q^r_m=\frac{e(\a_r)}{\sqrt{|\a_r|}}\bigl[C\overline{N}^r\bigr]_m\,.
\end{equation}
Taking into account \eqref{oscrel} we see that $|E_b\ra$ precisely
coincides with the flat space expression equation (4.21) of
\cite{gsb}.
\section{The constituents of the Prefactor}\label{sec5}
The full expression for the superstring vertex involves an
operator $G$, the so-called prefactor, necessary to ensure the
realization of the superalgebra in the interacting theory \cite{gs}. In flat space this
operator, when written in the continuum basis depends on
$p_r(\s)$, $\grave{x}_r(\s)$ and $\l_r(\s)$. Since $p_r(\s)$ and
$\l_r(\s)$ correspond to functional derivatives with respect to $x_r(\s)$
and $\vt_r(\s)$ the only physically sensible value of $\s$ to
choose is the interaction point $\s=\pm\pi\a_1$. Since operators
at this point are singular the prefactor must be carefully defined
in the limit $\s\to|\pi\a_1|$ \cite{gs}. In the end one obtains an
expression containing both creation and annihilation operators of
the various oscillators. The annihilation operators can be
eliminated by (anti)commuting them through the exponential factors
of the vertex. We then write $GE_aE_b|0\ra=E_aE_b\tilde{G}|0\ra$
where $\wt{G}$ only contains the creation operators.

In this section we study the bosonic and fermionic constituents of the
prefactor, relate the infinite component vectors that appear in their oscillator basis
expressions and point out a subtlety in the relation between the continuum and oscillator
basis expressions which is not present in the flat space case. This might have the
consequence that the precise form of the prefactor derived in \cite{sv} has to be changed.

\subsection{The bosonic constituents}
An important constraint on the prefactor is that it must respect
the local conservation laws ensured by $|E_a\ra$ and $|E_b\ra$.
For the bosonic part this means that it must commute with \cite{gs,gsb}
\begin{equation}\label{bospre}
\bigl[\,\sum_{r=1}^3p_r(\s),\wt{G}\bigr]=0=\bigl[\,\sum_{r=1}^3e(\a_r)x_r(\s),\wt{G}\bigr]\,.
\end{equation}
Consider first an expression of the form
\begin{equation}
{\mc K}_0+{\mc
K}_+=\sum_{r=1}^3\sum_{m=0}^{\infty}F_{m(r)}a^{\dg}_{m(r)}\,.
\end{equation}
The Fourier transform of \eqref{bospre} leads to the equations \cite{sv2}
\begin{equation}\label{kplus}
\sum_{r=1}^3\bigl[X^{(r)}C_{(r)}^{1/2}F_{(r)}\bigr]_m=0=\sum_{r=1}^3\a_r\bigl[X^{(r)}C_{(r)}^{-1/2}F_{(r)}\bigr]_m\,.
\end{equation}
Here the components $m=0$ and $m>0$ decouple from each other. For
$m=0$ one finds \cite{sv2}
\begin{equation}\label{f0}
F_{0(1)}=-\sqrt{\frac{2}{\a'}}\sqrt{\m\a_1}\a_2\,,\qquad
F_{0(2)}=-\sqrt{\frac{\a_1}{\a_2}}F_{0(1)}\,, \qquad F_{0(3)}=0\,.
\end{equation}
Then ${\mc K}_0$ can be written as
\begin{equation}
{\mc K}_0=\mathbb{P}-i\frac{\m\a}{\a'}\mathbb{R}
\end{equation}
which has the correct flat space limit. So in contrast to a statement in \cite{sv} the
term $\m\mathbb{R}$ is included in the supergravity part of the prefactor and although in
supergravity $\mathbb{R}|V\ra=0$ we have ${\mc K}_0^I{\mc K}_0^J|V\ra=
\bigl[\mathbb{P}^I\mathbb{P}^J+\frac{\m\a}{\a'}\d^{IJ}\bigr]|V\ra$ so the inclusion of
$\mathbb{R}$ does make a difference.\footnote{Moreover $\mathbb{R}|V\ra$ is no
longer zero in the full string theory, so to obtain a consistent expression for the
prefactor $\mathbb{R}$ has to be included, as was in fact done in \cite{sv2} when working
with the oscillator expressions.}
The overall normalization
of $F_{0(1)}$ is not determined by \eqref{kplus} and could be any
function $f(\m)$ with $\lim\limits_{\m\to0}f(\m)=1$.\footnote{In \cite{kim} this
normalization was fixed to be a constant by comparing the results of
\cite{sv} with a supergravity calculation.}
For $m>0$ we have
\begin{equation}\label{kplus2}
\sum_{r=1}^3\bigl[A^{(r)}C^{-1/2}C_{(r)}^{1/2}F_{(r)}\bigr]_m=\frac{1}{\sqrt{\a'}}\m\a
B_m=
\sum_{r=1}^3\m\a_r\bigl[A^{(r)}C^{-1/2}C_{(r)}^{-1/2}F_{(r)}\bigr]\,.
\end{equation}
Subtracting the second equation from the first one
\begin{equation}\label{f+1}
\sum_{r=1}^3\bigl[A^{(r)}C^{1/2}C_{(r)}^{-1/2}U_{(r)}F_{(r)}\bigr]_m=0\,.
\end{equation}
Using the first identity in \eqref{id2} gives \cite{sv2}
\begin{equation}
F_{m(r)}=\frac{1}{\a_r}\bigl[C_{(r)}^{1/2}C^{1/2}U_{(r)}^{-1}A^{(r)\,T}V\bigr]_m
\end{equation}
with $V_m$ an arbitrary vector determined by plugging the above
expression for $F_{m(r)}$ in, say, the second equation in
\eqref{kplus2}. The complete solution is \cite{sv2}
\begin{equation}
F_{m(r)}=\frac{\a}{\sqrt{\a'}}\frac{1}{\a_r}\bigl[C_{(r)}^{1/2}C^{1/2}U_{(r)}^{-1}A^{(r)\,T}\Upsilon^{-1}B\bigr]_m\,.
\end{equation}
The matrix $\Upsilon$ was introduced in section \ref{sec4}.
Using equation \eqref{ups2} one can show that
\begin{equation}\label{f+}
F_{m(r)}=-\frac{1}{\sqrt{\a'}}\frac{\a}{1-4\m\a
K}\frac{1}{\a_r}\bigl[U_{(r)}^{-1}C_{(r)}^{1/2}C\overline{N}^r\bigr]_m
\end{equation}
and as $\m\to0$
\begin{equation}
\lim_{\m\to0}\bigl({\mc K}_0+{\mc
K}_+\bigr)=\mathbb{P}-\frac{\a}{\sqrt{\a'}}\sum_{r=1}^3\sum_{m=1}^{\infty}\frac{1}{\a_r}
\bigl[C\overline{N}^r\bigr]_m\sqrt{m}a_{m(r)}^{\dg}
\end{equation}
coincides with the flat space result of \cite{gsb}.

From the expression for $F_{m(r)}$ in \eqref{f+} we see that the
equations in \eqref{kplus} are actually constraints on
$\overline{N}^r$
\begin{equation}\label{nr}
B+\sum_{r=1}^3A^{(r)}C^{1/2}U_{(r)}\overline{N}^r=0\,,\qquad
\sum_{r=1}^3\frac{1}{\a_r}A^{(r)}C^{3/2}\overline{N}^r=0
\end{equation}
precisely satisfied by
$\overline{N}^r=-C^{-1/2}A^{(r)\,T}\G^{-1}B$.

Now take into account the negative modes, i.e. consider
\begin{equation}
{\mc
K}_-=\sum_{r=1}^3\sum_{m=1}^{\infty}F_{-m(r)}a^{\dg}_{-m(r)}\,.
\end{equation}
This leads to the equations \cite{sv2}
\begin{equation}
\sum_{r=1}^3\frac{1}{\a_r}\bigl[A^{(r)}C^{1/2}C_{(r)}^{1/2}F_{(r)}\bigr]_{-m}=0=
\sum_{r=1}^3\bigl[A^{(r)}C^{1/2}C_{(r)}^{-1/2}F_{(r)}\bigr]_{-m}\,.
\end{equation}
Comparing the second equation with \eqref{f+1} it is clear that
\begin{equation}\label{f-+1}
F_{-m(r)}\sim U_{m(r)}F_{m(r)}\,.
\end{equation}
However, if one substitutes this into the first equation one actually sees that the sum is
divergent \cite{gs,gsb,sv2}. This phenomenon already appears in flat space and it is
known \cite{gs} that the function of $\s$ responsible for the divergence is
$\d(\s-\pi\a_1)-\d(\s+\pi\a_1)$. Since $\pm\pi\a_1$ are actually
identified this divergence is harmless but the precise normalization in \eqref{f-+1}
has to be determined by other means.

In flat space the prefactor can be alternatively defined as
follows. Consider the operators defined via their action on
$|V\ra$ \cite{gs,gsb}
\begin{equation}
\begin{split}
\p X|V\ra & =
4\pi\frac{\sqrt{-\a}}{\a'}\lim_{\s\to\pi\a_1}(\pi\a_1-\s)^{1/2}
\bigl(\grave{x}_1(\s)+\grave{x}_1(-\s)\bigr)|V\ra\,,\\
\label{p} P|V\ra &
=-2\pi\sqrt{-\a}\lim_{\s\to\pi\a_1}(\pi\a_1-\s)^{1/2}\bigl(p_1(\s)+p_1(-\s)\bigr)|V\ra\,.
\end{split}
\end{equation}
These expressions contain creation and annihilation operators. The
claim is that after commuting the annihilation operators through
the exponential and taking the limit $\s\to\pi\a_1$ we
have\footnote{Notice that in comparison with \cite{sv,sv2} we changed
the sign in the definition of $\p X$. Then ${\mc
K}\longleftrightarrow X_{\text{GS}}$ and $\wt{{\mc
K}}\longleftrightarrow \wt{X}_{\text{GS}}$ as compared to
\cite{gs}.}
\begin{equation}\label{preflat}
\begin{split}
\left(P+\frac{1}{4\pi}\p X\right)|V\ra & =\left({\mc K}_0+{\mc K}_++{\mc K}_-\right)|V\ra\equiv{\mc K}|V\ra\,,\\
\left(P-\frac{1}{4\pi}\p X\right)|V\ra & =\left({\mc K}_0+{\mc
K}_+-{\mc K}_-\right)|V\ra\equiv\wt{{\mc K}}|V\ra\,.
\end{split}
\end{equation}

\noindent Substituting the mode expansion of $p_1(\s)$ into \eqref{p} yields
\begin{align}
P|V\ra & \equiv\sum_{r=1}^3\sum_{m=0}^{\infty}P_{m(r)}a_{m(r)}^{\dg}|V\ra\nonumber\\
&=-\frac{2}{\a_1}\frac{\sqrt{-\a}}{\sqrt{\a'}}\lim_{e\to0}\e^{1/2}
\sum_{r=1}^3\sum_{m=0}^{\infty}\left[\sum_{n=1}^{\infty}(-1)^n\sqrt{\o_{n(1)}}\cos(n\e/\a_1)\overline{N}^{1r}_{nm}\right]
a_{m(r)}^{\dg}|V\ra\,.
\end{align}
Now the singular behavior of the sum as $\e\to0$ can be traced to
the way it diverges as $n\to\infty$. Therefore to take the limit
$\e\to0$ we can approximate the summand for large $n$. Thus for
$m>0$ and $n$ large
\begin{equation}
\begin{split}
\overline{N}^{1r}_{n0} & \sim -\sqrt{2\m\a_rn}\e^{rt}\a_t\overline{N}^1_n\,,\\
\overline{N}^{1r}_{nm} &\sim -\frac{\a}{\a_r}\frac{1}{1-4\m\a K}
\left(C^{1/2}\overline{N}^1\right)_n\left(U_{(r)}^{-1}C_{(r)}^{1/2}C\overline{N}^r\right)_m\,,
\end{split}
\end{equation}
where we used \eqref{nnpp}. Hence we find for $m>0$
\begin{equation}
P_{0(1)}=f(\m)F_{0(1)}\,,\qquad
P_{0(2)}=-\sqrt{\frac{\a_1}{\a_2}}P_{0(1)}\,,\qquad
P_{m(r)}=f(\m)F_{m(r)}\,.
\end{equation}
Here we defined
\begin{equation}
f(\m)\equiv-2\frac{\sqrt{-\a}}{\a_1}\lim_{e\to0}\e^{1/2}\sum_{n=1}^{\infty}(-1)^nn\cos(n\e/\a_1)\overline{N}^1_n
\end{equation}
which is equal to one for $\m=0$ \cite{gs}. It is not clear to us that or even if this
still holds for $\m$ non-zero but we will not need the precise form of $f(\m)$ in what
follows. Thus we have shown that $P|V\ra=f(\m)\bigl({\mc K}_0+{\mc K}_+\bigr)|V\ra$.
For the negative modes we get
\begin{align}
\frac{1}{4\pi}\p X|V\ra & \equiv\sum_{r=1}^3\sum_{m=1}^{\infty}\p X_{-m(r)}a^{\dg}_{-m(r)}|V\ra\nonumber\\
&=\frac{2i}{\sqrt{\a'}}\frac{\sqrt{-\a}}{\a_1}\lim_{\e\to0}\e^{1/2}
\sum_{r=1}^3\sum_{m=1}^{\infty}\left[\sum_{n=1}^{\infty}(-1)^n\frac{n}{\sqrt{\o_{n(1)}}}\cos(n\e/\a_1)
\overline{N}^{rs}_{-n,-m}\right]a_{-m(r)}^{\dg}|V\ra\,.
\end{align}
Using that
$\overline{N}^{1r}_{-n,-m}\sim-\overline{N}^{1r}_{nm}U_{m(r)}$ for large $n$ we
see that
\begin{equation}\label{f-+}
\p X_{-m(r)}=iU_{m(r)}P_{m(r)}\,.
\end{equation}
Hence $F_{-m(r)}=iU_{m(r)}F_{m(r)}$ which differs by a factor of
$i$ from the result of \cite{sv2}. As $\m\to0$ it reduces to the
flat space result of \cite{gsb}.
So we have seen that the continuum basis operators and the oscillator expressions coincide
up to a possibly $\m$-dependent normalization. Since we do not fix the normalization
anyway this is not important. We will see however in the next subsection that the
situation is slightly different in the fermionic case.
\subsection{The fermionic constituent}
The fermionic part of the prefactor  has to satisfy the conditions
\begin{equation}\label{ferpre}
\bigl\{\,\sum_{r=1}^3\l_r(\s),\wt{G}\bigr\}=0=\bigl\{\,\sum_{r=1}^3e(\a_r)\vt_r(\s),\wt{G}\bigr\}\,.
\end{equation}
Consider
\begin{equation}
{\mc
Y}=\sum_{r=1}^2G_{0(r)}\l_{0(r)}+\sum_{r=1}^3\sum_{m=1}^{\infty}G_{m(r)}b^{\dg}_{m(r)}\,.
\end{equation}
For the zero-modes we can set the coefficient of, say, $\l_{0(3)}$
to zero due to the property of the fermionic supergravity vertex that
$\sum_{r=1}^3\l_{0(r)}|E_b^0\ra=0$ . The
Fourier transform of \eqref{ferpre} leads to the equations
\begin{equation}\label{y}
\begin{split}
& \sum_{r=1}^3\frac{1}{\sqrt{|\a_r|}}\bigl[A^{(r)}CC_{(r)}^{-1/2}P_{(r)}G_{(r)}\bigr]_m=0\,,\\
& -\bigl[C^{1/2}B\bigr]_m\sum_{r,s=1}^2\e^{rs}\a_r\a_sG_{0(r)}+
\sum_{r=1}^3e(\a_r)\sqrt{|\a_r|}\bigl[C^{1/2}A^{(r)}C_{(r)}^{-1/2}P_{(r)}^{-1}G_{(r)}\bigr]_m=0\,.
\end{split}
\end{equation}
The components $m=0$ and $m>0$ decouple from each other (the term proportional to $B$
is absent for $m=0$). For $m=0$ we have
\begin{equation}\label{g0}
G_{0(1)}=-\sqrt{\frac{2}{\a'}}\a_2\,,\qquad
G_{0(2)}=\sqrt{\frac{2}{\a'}}\a_1\,.
\end{equation}
As in the previous subsection the normalization is only determined
up to a matrix $y_{ab}(\m)$ with $\lim\limits_{\m\to
0}y_{ab}(\m)=\d_{ab}$. For $m>0$ we can rewrite the second equation as
\begin{equation}
\sum_{r=1}^3e(\a_r)\sqrt{|\a_r|}\bigl[A^{(r)}C_{(r)}^{-1/2}P_{(r)}^{-1}G_{(r)}\bigr]_m=\frac{\a}{\sqrt{\a'}}B_m\,.
\end{equation}
Then
\begin{equation}
G_{m(r)}=\frac{e(\a_r)}{\sqrt{|\a_r|}}\bigl[P_{(r)}^{-1}C_{(r)}^{1/2}A^{(r)\,T}W\bigr]_m
\end{equation}
solves the first equation using \eqref{id2} with $W_m$ an
arbitrary vector that is determined by the second equation. The
final solution is
\begin{equation}
G_{m(r)}=\frac{\a}{\sqrt{a'}}\frac{e(\a_r)}{\sqrt{|\a_r|}}
\bigl[P_{(r)}^{-1}C_{(r)}^{1/2}A^{(r)\,T}\wt{\Upsilon}^{-1}B\bigr]_m\,.
\end{equation}
Here
\begin{equation}
\wt{\Upsilon}\equiv\sum_{r=1}^3A^{(r)}P_{(r)}^{-2}A^{(r)\,T}\,,\qquad
\wt{\Upsilon}^{-1}=\G^{-1}-\frac{1}{2}\frac{\m\a}{1-4\m\a K}
\left(\G^{-1}B\right)\left(\G^{-1}B\right)^T(1+\Pi)\,.
\end{equation}
Hence $G_{(r)}$ can be expressed via $F_{(r)}$ as
\begin{equation}\label{gf}
G_{(r)}=\bigl(1-2\m\a
K(1-\Pi)\bigr)\sqrt{|\a_r|}P_{(r)}^{-1}U_{(r)}C^{-1/2}F_{(r)}\,.
\end{equation}
As $\m\to0$ we have
\begin{equation}
\lim_{\m\to0}{\mc Y}=\sqrt{\frac{2}{\a'}}\L+
\sum_{r=1}^3\sum_{m=1}^{\infty}\frac{F_{m(r)}}{\sqrt{m}}\sqrt{|\a_r|}b_{m(r)}^{\dg}\,.
\end{equation}
Taking into account that
$\sqrt{|\a_r|}b_{m(r)}^{\dg}\longleftrightarrow Q_{-m(r)}^I$ this
is exactly the flat space expression of \cite{gsb}.
As a further check of the previous equations, in particular
\eqref{gf}, we note that upon substitution of $G_{(r)}$ one can
show that the constraints determining $G_{(r)}$ reduce again to
the two equations given in \eqref{nr}.
In flat space the operator
\begin{equation}
Y(\s)=-2\pi\frac{\sqrt{-2\a}}{\sqrt{\a'}}(\pi\a_1-\s)^{1/2}
\bigl(\l_1(\s)+\l_1(-\s)\bigr)
\end{equation}
satisfies $\lim\limits_{\s\to\pi\a_1}Y(\s)|V\ra={\mc Y}|V\ra$.
Substituting the mode expansion for $\l_1(\s)$ we get
\begin{equation}
Y|V\ra=-\sqrt{\frac{2}{\a'}}\sqrt{\frac{-2\a}{\a_1}}\lim_{\e\to0}\e^{1/2}
\sum_{n=1}^{\infty}(-1)^n\cos(n\e/\a_1)\left[\sqrt{2}\L Q^1_n+
\sum_{r=1}^3\sum_{m=1}^{\infty}Q^{1r}_{nm}b^{\dg}_{m(r)}
\right]|V\ra\,.
\end{equation}
For large $n$
\begin{equation}
\begin{split}
Q^1_n & \sim \frac{1}{\sqrt{\a_1}}(1-4\m\a K)^{-1}(1-2\m\a K(1+\Pi))\bigl[C\overline{N}^1\bigr]_n\,,\\
Q^{1r}_{nm} & \sim \sqrt{\frac{\a'}{\a_1}}(1-4\m\a K)^{-1}(1-2\m\a
K(1+\Pi)) \bigl[C\overline{N}^1\bigr]_nG_{m(r)}\,.
\end{split}
\end{equation}
Then
\begin{equation}
\begin{split}
Y_0 & = f(\m)(1-4\m\a K)^{-1}(1-2\m\a K(1+\Pi))\sqrt{\frac{2}{\a'}}\L\,,\\
Y_{m(r)} & = f(\m)(1-4\m\a K)^{-1}(1-2\m\a K(1+\Pi))G_{m(r)}\,.
\end{split}
\end{equation}
So we have shown that $Y|V\ra=f(\m)(1-4\m\a K)^{-1}(1-2\m\a
K(1+\Pi)){\mc Y}|V\ra$, i.e. for the fermionic constituent a non-trivial
$\m$-dependent matrix appears in the relative normalization.
\section{Conclusions}\label{sec6}
In this paper we studied light-cone superstring field theory on the maximally
supersymmetric pp-wave background. Our main results are a modified expression for the
fermionic contribution to the exponential part of the vertex as compared to \cite{sv},
the generalization of a flat space factorization theorem \cite{gs} to the pp-wave
background (see also \cite{sch}) and simple expressions for the fermionic Neumann
matrices in terms of the bosonic ones. We analyzed the oscillator and continuum basis
expressions for the constituents of the prefactor in detail and pointed out that in
contrast to flat space the relation between the two is non-trivial in the fermionic case.
As a consistency check, we have shown that all expressions appearing in the exponential as
well as the prefactor part of the interaction vertex coincide with the flat space results
of \cite{gsb} in the limit $\m\to0$.
\vskip1cm
\section*{Acknowledgment}
I am grateful for discussions with Gleb Arutyunov and Jan Plefka. In particular I 
would like to thank Bogdan Stefanski and Stefan Theisen for very useful comments
on the manuscript. This work was supported by GIF, the German-Israeli foundation for
Scientific Research.
\appendix
\section{Some summation formulae}
\setcounter{section}{1}
\renewcommand{\thesection}{\Alph{section}}
\setcounter{equation}{0}
It is convenient to introduce the matrices for $m,n>0$
\begin{equation}
\begin{split}
C_{mn} & = m\d_{mn}\,,\\
A^{(1)}_{mn} & = (-1)^n\frac{2\sqrt{mn}\b}{\pi}\frac{\sin
m\pi\b}{m^2\b^2-n^2}
=\bigl(C^{-1/2}X^{(1)}C^{1/2}\bigr)_{mn}\,,\\
A^{(2)}_{mn} & = \frac{2\sqrt{mn}(\b+1)}{\pi}\frac{\sin
m\pi\b}{m^2(\b+1)^2-n^2}
=\bigl(C^{-1/2}X^{(2)}C^{1/2}\bigr)_{mn}\,,\\
A^{(3)}_{mn} & = \d_{mn}
\end{split}
\end{equation}
and the vector $(m>0)$
\begin{equation}
B_m=-\frac{2}{\pi}\frac{\a_3}{\a_1\a_2}m^{-3/2}\sin m\pi\b
\end{equation}
related to $X^{(r)}_{m0}$ via
\begin{equation}
X^{(r)}_{m0}=-\e^{rs}\a_s\bigl(C^{1/2}B\bigr)_m\,.
\end{equation}
These quantities satisfy some very important identities that were derived in
\cite{gs}. They are
\begin{equation}\label{id1}
-\frac{\a_3}{\a_r}\,CA^{(r)T}C^{-1}A^{(s)}=\d^{rs}{\bf 1}\,,\quad
-\frac{\a_r}{\a_3}\,C^{-1}A^{(r)T}CA^{(s)}=\d^{rs}{\bf 1}\,,\quad
A^{(r)T}CB=0
\end{equation}
for $r,s\in\{1,2\}$ and
\begin{equation}\label{id2}
\sum_{r=1}^3\frac{1}{\a_r}A^{(r)}CA^{(r)\,T}=0\,,\qquad
\sum_{r=1}^3\a_rA^{(r)}C^{-1}A^{(r)\,T}=\frac{\a}{2}BB^T\,,
\end{equation}
where $\a\equiv\a_1\a_2\a_3$. In terms of the big matrices
$X^{(r)}_{mn}$, $m,n\in{\bf Z}$ the relations \eqref{id1} and
\eqref{id2} can be written in the compact form
\begin{equation}
\bigl(X^{(r)T}X^{(s)}\bigr)_{mn}=-\frac{\a_3}{\a_r}\d^{rs}\d_{mn}\,,\quad
r,s\in\{1,2\}\,,\qquad
\sum_{r=1}^3\a_r\bigl(X^{(r)}X^{(r)T}\bigr)_{mn}=0\,.
\end{equation}
\section{The exponential part of the vertex}\label{app:b}
\setcounter{section}{2}
\renewcommand{\thesection}{\Alph{section}}
\setcounter{equation}{0}
\subsection*{The bosonic part}
The bosonic constraints the exponential part of the vertex has to
satisfy are
\begin{equation}
\sum_{r=1}^3\sum_{n\in{\bf Z}}X_{mn}^{(r)}p_{n(r)}|V\ra=0\,,\qquad
\sum_{r=1}^3\sum_{n\in{\bf Z}}\a_rX_{mn}^{(r)}x_{n(r)}|V\ra=0\,.
\end{equation}
For $m=0$ this leads to
\begin{equation}
\sum_{r=1}^3p_{0(r)}|V\ra=0\,,\qquad
\sum_{r=1}^3\a_rx_{0(r)}|V\ra=0\,.
\end{equation}
Substituting \eqref{xp} and commuting the annihilation operators
through the exponential this requires
\begin{align}
&\sum_{r,s=1}^3\sqrt{|\a_r|}\bigl[\bigl(\overline{N}^{rs}_{00}+\d^{rs}\bigr)a_{0(s)}^{\dg}
+\sum_{n=1}^{\infty}\overline{N}^{rs}_{0n}a_{n(s)}^{\dg}\bigr]|V\ra=0\,,\\
&\sum_{r,s=1}^3e(\a_r)\sqrt{|\a_r|}\bigl[\bigl(\overline{N}^{rs}_{00}-\d^{rs}\bigr)a_{0(s)}^{\dg}
+\sum_{n=1}^{\infty}\overline{N}^{rs}_{0n}a_{n(s)}^{\dg}\bigr]|V\ra=0\,.
\end{align}
Using the expressions given for $\overline{N}^{rs}_{0n}$ and
$\overline{N}^{rs}_{00}$ in \eqref{m0}, \eqref{00a} and
\eqref{00b} one can check that the above equations are satisfied
trivially, i.e. without further use of additional non-trivial
identities. For $m>0$ we find the following constraints
\begin{align}
\label{bos1}
B+\sum_{r=1}^3A^{(r)}C^{1/2}U_{(r)}\overline{N}^r & = 0\,,\\
\label{bos2}
A^{(s)}C_{(s)}^{-1/2}U_{(s)}^{-1}+\sum_{r=1}^3A^{(r)}C_{(r)}^{-1/2}U_{(r)}C^{1/2}\overline{N}^{rs}C^{-1/2} & = 0\,,\\
\label{bos3}
-\a_sA^{(s)}C_{(s)}^{-1/2}+\sum_{r=1}^3\a_rA^{(r)}C_{(r)}^{-1/2}C^{-1/2}\overline{N}^{rs}C^{1/2}
& = \a B\bigl[C_{(s)}^{1/2}C^{1/2}\overline{N}^s\bigr]^T\,.
\end{align}
Equation \eqref{bos1} is identical to the first equation in
\eqref{nr}. To prove equations \eqref{bos2} and \eqref{bos3}
substitute the expression for $\overline{N}^{rs}$ given in
\eqref{mn}.
For $m<0$ there is one additional constraint
\begin{equation}
A^{(s)}C_{(s)}^{-1/2}U_{(s)}^{-1}-
\a_s\sum_{r=1}^3\frac{1}{\a_r}A^{(r)}C_{(r)}^{1/2}U_{(r)}C^{1/2}\overline{N}^{rs}C^{-1/2}C_{(s)}^{-1}=0
\end{equation}
which can be verified by subtracting it from equation \eqref{bos2} and
using \eqref{nnpp2}. Here the identity
\begin{equation}
\sum_{r=1}^3\a_rA^{(r)}C^{-1/2}\overline{N}^r=2\a K B\,.
\end{equation}
is used.
\subsection*{The fermionic part}
The fermionic constraints the exponential part of the vertex has
to satisfy are
\begin{equation}\label{lt}
\sum_{r=1}^3\sum_{n\in{\bf
Z}}X_{mn}^{(r)}\l_{n(r)}|V\ra=0\,,\qquad
\sum_{r=1}^3\sum_{n\in{\bf Z}}\a_rX_{mn}^{(r)}\vt_{n(r)}|V\ra=0\,.
\end{equation}
For $m=0$ this leads to
\begin{equation}
\sum_{r=1}^3\l_{0(r)}|V\ra=0\,,\qquad
\sum_{r=1}^3\a_r\vt_{0(r)}|V\ra=0\,.
\end{equation}
These equations are satisfied by construction of the zero-mode
part of $|V\ra$.
For $m>0$ we get
\begin{align}
\label{ferm1}
B+\sum_{r=1}^3e(\a_r)\sqrt{|\a_r|}A^{(r)}C_{(r)}^{-1/2}P_{(r)}Q^r & = 0\,,\\
\label{ferm2} \sqrt{|\a_s|}A^{(s)}C_{(s)}^{-1/2}P_{(s)}^{-1}+
\sum_{r=1}^3e(\a_r)\sqrt{|\a_r|}A^{(r)}C_{(r)}^{-1/2}P_{(r)}Q^{rs} & = 0\,,\\
\label{ferm3}
-\sqrt{|\a_s|}A^{(s)}C_{(s)}^{-1/2}P_{(s)}+
\sum_{r=1}^3e(\a_r)\sqrt{|\a_r|}A^{(r)}C_{(r)}^{-1/2}P_{(r)}^{-1}Q^{sr\,T}
& = \a BQ^{s\,T}
\end{align}
and $\a_3\Theta|V_0\ra\equiv\bigl(\vt_{0(1)}-\vt_{0(2)}\bigr)|V_0\ra=0$. The latter equation
is satisfied by the zero-mode part of the vertex given in equation \eqref{zero}.
For $m<0$ the constraints are
\begin{align}
\label{ferm4}
\sum_{r=1}^3\frac{1}{\sqrt{|\a_r|}}A^{(r)}CC_{(r)}^{-1/2}P_{(r)}^{-1}Q^r & = 0\,,\\
\label{ferm5}
A^{(s)}CC_{(s)}^{-1/2}-e(\a_s)\sqrt{|\a_s|}\sum_{r=1}^3\frac{1}{\sqrt{|\a_r|}}
A^{(r)}CC_{(r)}^{-1/2}P_{(r)}^{-1}Q^{rs}P_{(s)}^{-1} & = 0\,,\\
\label{ferm6}
A^{(s)}CC_{(s)}^{-1/2}+e(\a_s)\sqrt{|\a_s|}\sum_{r=1}^3\frac{1}{\sqrt{|\a_r|}}
A^{(r)}CC_{(r)}^{-1/2}P_{(r)}Q^{sr\,T}P_{(s)} & = 0\,.
\end{align}
Now equations \eqref{ferm1} and \eqref{ferm4} uniquely determine
\begin{equation}
Q^r=\frac{e(\a_r)}{\sqrt{|\a_r|}}(1-4\m\a K)^{-1}(1-2\m\a
K(1+\Pi))P_{(r)}C_{(r)}^{1/2}C^{1/2}\overline{N}^r\,.
\end{equation}
Furthermore comparing equations \eqref{ferm2} and \eqref{bos2} we
see that
\begin{equation}
Q^{rs}=e(\a_r)\sqrt{\left|\frac{\a_s}{\a_r}\right|}
P_{(r)}^{-1}U_{(r)}C^{1/2}\overline{N}^{rs}C^{-1/2}U_{(s)}P_{(s)}^{-1}
\end{equation}
solves \eqref{ferm2}. Using
\begin{equation}
P^{-2}_{(r)}U_{(r)}\overline{N}^{rs}U_{(s)}P_{(s)}^{-2}=\overline{N}^{rs}
+\m\a(1-4\m\a
K)^{-1}C_{(r)}^{1/2}\overline{N}^r\bigl[C_{(s)}^{1/2}\overline{N}^s\bigr]^T(1-\Pi)
\end{equation}
establishes \eqref{ferm3} by virtue of \eqref{bos3}.
Equation \eqref{ferm5} is satisfied due to the identity
\begin{equation}
A^{(s)}C_{(s)}^{-1/2}-\a_s\sum_{r=1}^3\frac{1}{\a_r}A^{(r)}C_{(r)}^{-1/2}
C^{3/2}\overline{N}^{rs}C^{-3/2}=0
\end{equation}
which can be proved using the expression for $\overline{N}^{rs}$ given in \eqref{mn}.
Finally, equation \eqref{ferm6} is identical to \eqref{bos2}. This
concludes the determination of the exponential part of the fermionic vertex.
\section{A note on the fermionic zero-mode vertex}\label{app:c}
\setcounter{section}{3}
\renewcommand{\thesection}{\Alph{section}}
\setcounter{equation}{0}
Here we show that the recent proposal of~\cite{chu1} for the
fermionic zero-mode vertex can be extended to a solution of the full kinematic constraints eq.~(\ref{lt}), see 
also~\cite{chu2}.\footnote{In a 
previous version 
of this paper it was erroneously claimed that this extension is not possible. I would like to thank the authors of~\cite{chu2} for 
useful correspondence about this point.}
Recall that the proposal of~\cite{chu1} is to use
\begin{equation}\label{delta1}
|\d\ra=\prod_{a=1}^8\left[\sum_{r,s=1}^3\a_s\l_{0(r)}^a
\vt_{0(s)}^a\right]|v\ra
\end{equation}
instead of $|E_b^0\ra$ as the fermionic zero-mode vertex. Here
$|v\ra$ is the pp-wave vacuum that is annihilated by all the $b_{n(r)}$.
Expressing $\l_{0(r)}$ and $\vt_{0(r)}$ in
terms of $b_{0(r)}$, $b_{0(r)}^{\dg}$ it is easy to see that
\begin{equation}\label{delta2}
|\d\ra=\a_3^8\prod_{a=1}^8
\left[1-\left(\sqrt{-\b}\,b_{0(1)}^{\dg}+
\sqrt{\b+1}\,b_{0(2)}^{\dg}\right)b_{0(3)}^{\dg}\right]|v\ra\,.
\end{equation}
Now the most general ansatz for the exponential part of the fermionic vertex is
\begin{equation}
|E_b\ra\sim\exp\left[\sum_{r,s=1}^3\sum_{m,n=0}^{\infty}b_{-m(r)}^{\dg}Q^{rs}_{mn}b_{n(s)}^{\dg}\right]|v\ra\,.
\end{equation}
Using the identities of appendix~\ref{app:b} it was shown in~\cite{chu2} that the kinematic constraints eq.~(\ref{lt}) 
are solved by 
\begin{eqnarray}
Q^{rs}_{mn} && = \frac{1+\Pi}{2}\bar{Q}^{rs}_{mn}-\frac{1-\Pi}{2}\bar{Q}^{sr}_{nm}
=\frac{1+\Pi}{2}\bar{Q}^{rs}_{mn}
-\frac{1-\Pi}{2}\frac{\a_r}{\a_s}\bigl(C^{-1}\bar{Q}^{rs}C\bigr)_{mn}\,,\\
Q^{rs}_{m0} && = -\frac{1+\Pi}{2}\e^{st}\sqrt{\a_s}\a_t\bar{Q}^r_m\,,\\
Q^{rs}_{0n} && = \frac{1-\Pi}{2}\e^{rt}\sqrt{\a_r}\a_t\bar{Q}^s_n\,,\\
Q^{3r}_{00} && = -Q^{r3}_{00} = \frac{1}{2}\sqrt{-\frac{\a_r}{\a_3}}\,,
\end{eqnarray}
and in terms of the bosonic Neumann matrices we have
\begin{eqnarray}
\bar{Q}^{rs} && = e(\a_r)\sqrt{\left|\frac{\a_s}{\a_r}\right|}
U_{(r)}^{1/2}C^{1/2}\bar{N}^{rs}C^{-1/2}U_{(s)}^{1/2}\,,\\
\bar{Q}^r && = \frac{e(\a_r)}{\sqrt{|\a_r|}}\bigl(U_{(r)}C_{(r)}C\bigr)^{1/2}\bar{N}^r\,.
\end{eqnarray}
To compare with~\cite{chu2} note that
\begin{eqnarray}
\frac{1+\Pi}{2}\sum_{s,t}\e^{st}\sqrt{\a_s}\a_tb_{0(s)}^{\dg}
&& =-\frac{1+\Pi}{2}\sqrt{2}\L\,,\\
\frac{1-\Pi}{2}\sum_{s,t}\e^{st}\sqrt{\a_s}\a_tb_{0(s)}^{\dg}
&& =\frac{1-\Pi}{2}\frac{\a}{\sqrt{2}}\Theta\,.
\end{eqnarray}

\end{document}